\documentclass[prl,twocolumn,superscriptaddress,appendix]{revtex4}
\usepackage{graphicx,amsmath}

\newcommand{\beq}{\begin{equation}}
\newcommand{\eeq}{\end{equation}}
\newcommand{\beqa}{\begin{eqnarray}}
\newcommand{\eeqa}{\end{eqnarray}}

\def\half{\frac{1}{2}}

\def\half{\frac{1}{2}}

\def\<{\langle}
\def\>{\rangle}
\def\ket#1{|#1\rangle}

\newcommand{\complex}{{\kern .1em {\raise .47ex\hbox {$\scriptscriptstyle |$}}\kern -.4em {\rm C}}}
\newcommand{\real}{{{\rm I} \kern -.19em {\rm R}}}
\def\opone{\leavevmode\hbox{\small1\normalsize\kern-.33em1}}

\begin{document}

\title{Proposal for Implementing Device-Independent Quantum Key Distribution\\
 based on a Heralded Qubit Amplification}

\author{Nicolas Gisin\footnotemark[1] \footnotetext{\footnotemark[1] The authors contributed equally
to this work.}}
\affiliation{
    Group of Applied Physics, University of Geneva, 1211 Geneva 4,
    Switzerland}

\author{Stefano Pironio\footnotemark[1]}
\affiliation{
    Group of Applied Physics, University of Geneva, 1211 Geneva 4,
    Switzerland}
\affiliation{
    Laboratoire d'Information Quantique, Universit\'e Libre de Bruxelles, Belgium}

\author{Nicolas Sangouard\footnotemark[1]}
\affiliation{
    Group of Applied Physics, University of Geneva, 1211 Geneva 4,
    Switzerland}

\date{\today}

\begin{abstract}
In device-independent quantum key distribution (DIQKD), the violation of a Bell inequality is exploited to establish a shared key that is secure independently of the internal workings of the QKD devices.
An experimental implementation of DIQKD, however, is still awaited, since hitherto all optical Bell tests are subject to the detection loophole, making the protocol unsecured. In particular, photon losses in the quantum channel represent a fundamental limitation for DIQKD. Here, we introduce a heralded qubit amplifier based on single-photon sources and linear optics that provides a realistic solution to overcome the problem of channel losses in Bell tests. \end{abstract}

\maketitle
%Quantum Key Distribution (QKD) enjoys an enviable position in physics, between fundamental quantum mechanics and applied quantum optics \cite{RMP}. From a fundamental point of view, all security proofs of QKD rely on the elegant theory of entanglement. To conclude about the presence of entanglement in a real system, however, one should know precisely what has been measured, e.g., which self-adjoint operators describe the arrangement of cables and lasers \cite{Acin06}. When QKD is performed in a lab, by well trained physicists, this may not be a problem. But the situation changes completely if QKD should operate obliviously between two black boxes installed, e.g., in banks. In practice, the solution consists in thoroughly testing the black QKD-boxes \cite{lo,makarov,scarani} and eventually to submit them to some certification agency. But there is a more elegant solution.

%A possible way to detect entanglement is to perform measurements whose output correlations violate Bell inequalities \cite{bell}.

Bell inequalities had an enormous impact on the foundations of quantum physics~\cite{bell}. Interestingly, they also find application in Device-Independent Quantum Key Distribution (DIQKD) \cite{Ekert91,mayers-yao,QKD-DI-PRL,QKD-DIFullLength,mckague,bhk,masanes}: as their violation guarantees the presence of entanglement independently of what precisely is measured, they can be exploited to establish a secret key between two black boxes without the necessity to know anything about how the boxes operate (see Figure~1). %The only requirement is that they do not communicate arbitrarily. This ``no communication" requirement is anyway a must for cryptographic boxes: everything that comes out of the boxes should be under control, if not they could merely send the secret to the adversary.  Note the revolution: testing Bell inequalities in no longer a quest on the foundations, but becomes part of applied physics.
\begin{figure}[t]
{\includegraphics[scale=0.60]{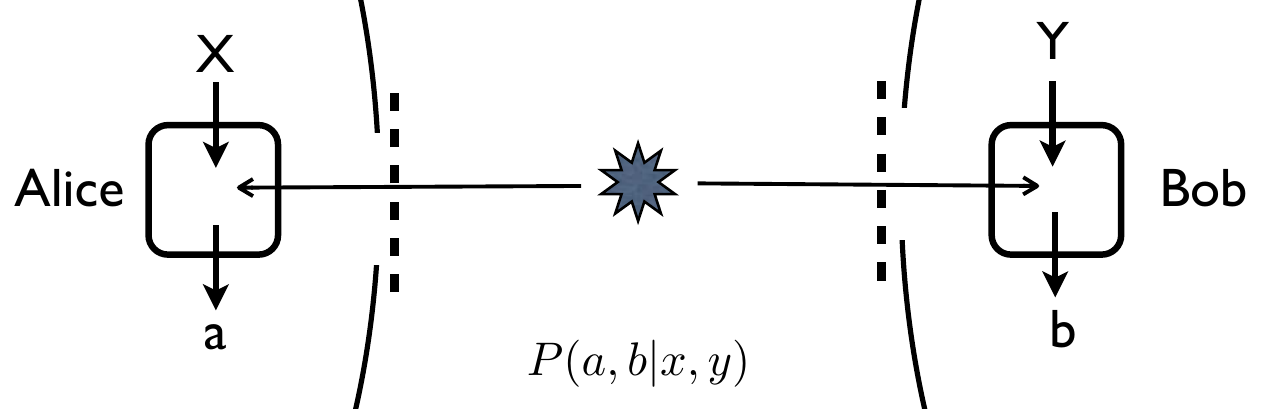}
 \caption{Principle of DIQKD.  Alice and Bob repeatedly choose inputs $x$ and $y$  for their QKD devices and obtain outputs $a$ and $b$. The inputs can be thought of as measurements on entangled particles, and the outputs as the measurement outcomes. At the end of the protocol, Alice and Bob use an authenticated public channel to compare a sample of their data in order to estimate the conditional probability distribution $P(a,b|x,y)$. If $P(a,b|x,y)$ violates the CHSH-Bell inequality by a sufficient amount, then Alice and Bob can use standard error correction and privacy amplification to distill a secret key out of the remaining data. To establish security nothing has to be known or assumed about Alice's and Bob's black boxes, except that they can be described by quantum physics.\\
Note, however, that it is assumed that Alice and Bob are each located in a secure place and control the information going in and out of their locations (dotted lines). In particular, the value of the inputs $x,y$ and of the outputs $a,b$ should not leak out unwillingly of Alice's and Bob's secure place. This is the only part of the protocol that cannot be untrusted: Alice and Bob should either enforce these conditions (e.g., by closing a ``door'') or test it (e.g., by monitoring the output signals of the boxes). %Note, however, that there is no way to establish confidential communication without assuming that Alice and Bob can indeed secure their places and that there is no information leakage: this is merely a basic assumption in all cryptographic scenario, whether quantum or not.
}
\label{fig1}}
\end{figure}

An experimental demonstration of DIQKD, however, is still awaited. Indeed, all optical tests of Bell's inequality suffer from the {\it detection loophole} \cite{PearleDetLoophole}: not all entangled photons are detected, because of unavoidable losses in the quantum channel, losses in the coupling between the photon-pair source and the optical fibers, and because of finite detector efficiency. The usual way out in Bell tests consists in assuming that the set of detected photon pairs is a fair set (the fair sampling assumption). It is indeed reasonable to assume that Nature is not malicious and does not trick us. But the situation is completely different in DIQKD. Here one does not test Nature, but fights against a possible active adversary \cite{lo,makarov}: it would make no sense to assume that the eavesdropper is not malicious.
Missed events could be used to perform simple and powerful attacks, e.g. the eavesdropper could force the black boxes to produce results only if the settings of the measuring devices are in agreement with a predeterminate scheme.
Closing the detection loophole in an optical experiment is therefore a requirement for a demonstration of DIQKD.

The detection efficiency, the product of the transmission efficiency (including the coupling into the fiber) and the photon-detector efficiency, required to rule out attacks based on the detection loophole is very high, typically larger than $82.8$\% for the CHSH inequality in the absence of other limitations. However, even assuming perfect photo-detection and lossless components, the transmission efficiency of a 5 km long optical-fiber at telecom wavelength is roughly of 80\%. Transmission losses thus represent a fundamental limitation for the realization of a detection-loophole free Bell test on any distance relevant for QKD.

The problem of transmission losses might be circumvented by performing quantum-non-demolition measurements of the incoming photon or by using quantum repeaters to distribute entanglement over large distances \cite{qr} in a heralded way. Here, we propose a much simpler scheme based on heralded qubit amplification that combines single-photon sources and linear optical elements only. Our proposal could be implemented with present-day technology. It provides a realistic avenue towards device-independent quantum cryptography.

\paragraph{Heralded qubit amplifier.}
Recently Ralph and Lund proposed a clever use of quantum teleportation to realize a heralded single-photon amplifier \cite{Ralph08}. Their scheme, presented in Fig. 2.a), has already motivated several experiments \cite{Pryde09, Grangier09}. We show how it can be extended for polarization-qubit amplification and we describe how this can be used in long-distance Bell experiments.

We consider a (normalized) coherent superposition
\beq
\psi_{in} =\alpha|0\rangle+\left(\beta_h in^\dagger_h +\beta_v in^\dagger_v\right)|0\rangle\nonumber
\eeq
of a vacuum component and of a qubit corresponding to a single photon either horizontally (corresponding to the creation operator $in_h^\dagger$) or vertically polarized (associated to $in_v^\dagger$). This state enters the device presented in Fig. 2.b). Two auxiliary photons, one horizontally $|1_h\rangle$ polarized and the other one vertically $|1_v\rangle$ polarized, are sent through a beamsplitter with transmission $t$. This leads to the entanglement $\left(\sqrt{1-t}c_h^\dagger + \sqrt{t}\-\ {out}_h^\dagger \right) \otimes \left(\sqrt{1-t}c_v^\dagger + \sqrt{t}\-\ {out}_v^\dagger \right)  |0\rangle$ of modes $c$ and $out$. The modes $c_{h,v}$ and $in_{h,v}$  are then combined on a $50/50$ beamsplitter. The modes after this beamsplitter are $d_h=(c_h+in_h)/\sqrt{2}$, $\tilde{d}_h=(c_h-in_h)/\sqrt{2}$, $d_v=(c_v+in_v)/\sqrt{2}$, and $\tilde{d}_v=(c_v-in_v)/\sqrt{2}$. The detection of two photons with orthogonal polarization, for example, one in mode $d_h$, the other one in mode $d_v$, projects the output mode into
\beq
\psi_{out}=\frac{\sqrt{1-t}}{2}\left(\sqrt{1-t}\alpha |0\rangle+\sqrt{t}\left(\beta_h {in}_h^\dagger+\beta_v {in}_v^\dagger\right)|0\rangle\right)\nonumber.
\eeq
For $t=1/2,$ the output state is equal to the input state and the scheme reduces to a teleportation protocol for qutrits with a partial Bell state analyzer. But for $t>1/2$, the relative weight of the vacuum component decreases, leading to the amplification of the polarization-qubit. This qubit amplification is probabilistic, since it depends on the accomplishment of the Bell measurement, but it is heralded by two detector clicks. The success probability is given by $|\psi_{out}|^2$. Since the detection of two-photons in modes ($d_h$,$\tilde{d}_v$), ($\tilde{d}_h$,$d_v$), or ($\tilde{d}_h$,$\tilde{d}_h$) combined with the appropriate one-qubit rotation also collapses the outcoming state into $\psi_{out},$ the overall success probability of the heralding amplifier is given by $4|\psi_{out}|^2.$
\begin{figure}
{
\includegraphics[scale=0.60]{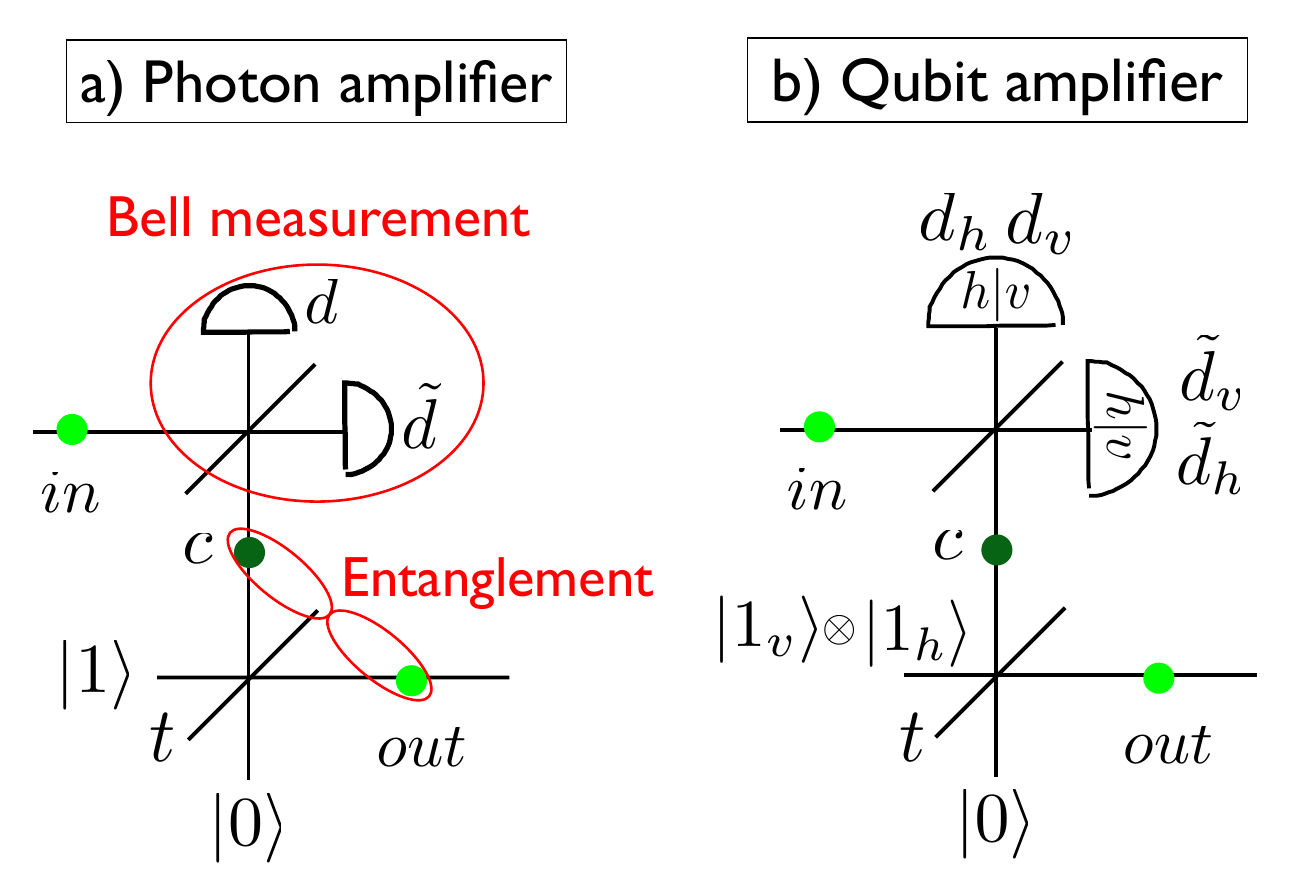}
\caption{a) Heralded amplifier for single photons as proposed in Ref. \cite{Ralph08}. A beam splitter with transmission coefficient $t$ turns an incoming photon into the entanglement of modes $c$ and ${out}$ which can be used to teleport an arbitrary state $\alpha \ket{0} + \beta {in}^\dagger \ket{0}$ with the help of a partial Bell state analyzer. If $t=\half$, this is standard quantum teleportation, i.e. the outcoming state $\alpha \ket{0} \pm \beta {out}^\dagger \ket{0}$  is similar to the incoming one, up to a possible unitary transformation depending on which detector clicked. But if $t>\half$, a successful Bell state measurement projects the outcoming state in the incoming one but shifted towards the single-photon state $\ket{1}$: $\sqrt{1-t}\alpha\ket{0}\pm \sqrt{t} \beta {in}^\dagger \ket{0}$. b) Setup for amplifying polarization qubits in a heralded way. This scheme is similar to the single-photon amplifier except that a product state of two photons with orthogonal polarization are sent through the partial beamsplitter. The probabilistic Bell measurement is based on a 50-50 beamsplitter followed by polarization measurements in the $h/v$ basis (which require a polarization beamsplitter and two photodetectors). For $t=\half,$ a successful Bell measurement teleports an arbitrary qutrit of the form $\alpha |0\rangle+(\beta_h {in}_h^\dagger+\beta_v {in}_v^\dagger)|0\rangle.$ For $t>1/2,$ the teleported state $\sqrt{1-t}\alpha |0\rangle+\sqrt{t}\left(\beta_h {in}_h^\dagger+\beta_v {in}_v^\dagger\right)|0\rangle$ has a smaller vacuum component leading to the heralded amplification of the qubit state.}
\label{fig2}}
\end{figure}

\paragraph{Application to DIQKD.}
As all teleportation protocols, the qubit amplifier also applies to mixed states. This provides a powerful tool to overcome the problem of losses in the frame of DIQKD. Suppose that a photon-pair source located on Alice's side is excited and can emit entangled photons with a small probability $p\ll1$, leading to the state
\beq
|0\rangle\langle0|+p|\frac{a_h^\dagger b_h^\dagger+a_v^\dagger b_v^\dagger}{\sqrt{2}}\rangle\langle \frac{a_h^\dagger b_h^\dagger+a_v^\dagger b_v^\dagger}{\sqrt{2}}| + O(p^2)\,.
\eeq
The term $O(p^2)$ introduces errors in the protocol, leading to the requirement that $p$ has to be kept small. The mode $b$ is sent to Bob through a quantum channel and because of losses, Alice and Bob share the state
\begin{multline}
|0\rangle\langle 0| +\frac{1}{2}p\left(1-\eta_t\right) \left(|a_h^\dagger\rangle\langle a_h^\dagger |+|a_v^\dagger\rangle\langle a_v^\dagger |\right)\allowdisplaybreaks\\
+{p\eta_t}|\frac{a_h^\dagger b_h^\dagger+a_v^\dagger b_v^\dagger}{\sqrt{2}}\rangle\langle \frac{a_h^\dagger b_h^\dagger+a_v^\dagger b_v^\dagger}{\sqrt{2}}|\,,
\end{multline}
where $\eta_t$ denotes the transmission efficiency of the quantum channel.
\begin{figure}
{\includegraphics[scale=0.65]{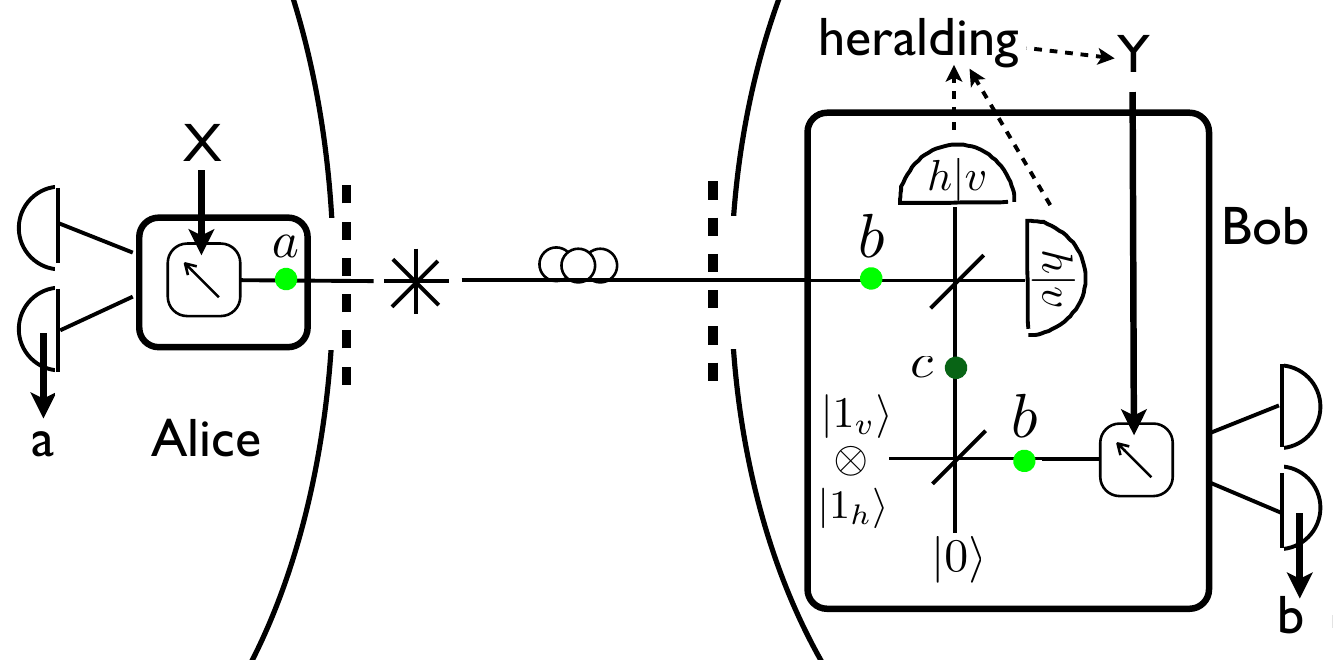}
 \caption{Proposed setup for the implementation of DIQKD based on a heralded qubit amplifier. The entangled-photon source is located close to Alice's location. Each of Alice's and Bob's black boxes includes a measurement apparatus. Furthermore, Bob's box contains the qubit amplifier which gives an heralding signal each time an entangled pair has been successfully distributed. Since Bob performs a measurement or, in other words, inputs a $y,$ only when he got the heralding signal, Alice and Bob can safely discard all events where a photon got lost in the quantum channel. Note that the detectors can be either out or in the boxes depending on whether they can be trusted or not. In the figure, they are outside the black boxes.}
\label{fig3}}
\end{figure}
Before Bob performs measurements, he amplifies the modes $b_h$ and $b_v$ using the setup described in Fig. \ref{fig3}. The state resulting from the successful amplification of both polarization modes is given by
\begin{eqnarray}
\label{ampli}
&&\frac{(1-t)^2}{4}|0\rangle\langle 0| \allowdisplaybreaks\\
&& \nonumber + \frac{(1-t)^2 p\left(1-\eta_t\right)}{8} \left(|a_h^\dagger\rangle\langle a_h^\dagger |+|a_v^\dagger\rangle\langle a_v^\dagger |\right)\allowdisplaybreaks\\
&&\nonumber + \frac{t (1-t)p\eta_t}{4} |\frac{a_h^\dagger b_h^\dagger+a_v^\dagger b_v^\dagger}{\sqrt{2}}\rangle\langle \frac{a_h^\dagger b_h^\dagger+a_v^\dagger b_v^\dagger}{\sqrt{2}}|.
\end{eqnarray}
For large enough $t,$ the entangled component is amplified in a heralded way, offering the possibility for Alice and Bob to share a maximally entangled state despite losses. This promises a considerable advance towards the implementation of DIQKD on meaningful distances. The heralding signal from the amplifier allows Bob to introduce an input $y$ in his black box only when he shares an entangled state with Alice. Hence, the overall detection efficiency required to close the detection loophole does not depend anymore on the transmission efficiency, but reduces to the intrinsic detection efficiency of Alice's and Bob's boxes.

The probability to obtain a heralded signal is
\beq\label{proba}
P_{H}=(1-t)^2+p(1-t)^2(1-\eta_t)+t(1-t)p\eta_t
\eeq
which roughly reduces to $(1-t)^2$ for small transmission efficiency. As can be seen from Eqs.~(\ref{ampli}) and (\ref{proba}), there is a tradeoff on the transmission coefficient $t$ of the partial beamsplitter. The amplification of the entangled component favors $t \approx 1,$ whereas a high success probability favors $t \approx 0.$ In order to rule out attacks based on the detection loophole, it is essential to choose a large transmission coefficient $t \approx 1$ to guarantee the distribution of highly entangled states. The price to pay is a reduction in the key rate because of the limited success probability of the qubit amplifier. Note that the problem of transmission losses cannot be overcome using a standard quantum relay implemented with two remote stochastic photon-pair sources and a Bell measurement made with linear optical elements and photon detectors. Indeed, without post-selection, a standard quantum relay allows only Alice and Bob to share a poorly entangled state due to multi-pair emissions.

\paragraph{Implementation and performance analysis.}
In practice, photons get lost not only because of the transmission losses in the quantum channel but also because of the imperfect coupling of photons into the optical fibers, which is characterized by an efficiency $\eta_c$. On Bob's side, the coupling loss can be counter-balanced by the amplifier, as the transmission losses. However, the amplifier itself contributes a factor $\eta_c$ back to the detection efficiency of Bob's box since the single-photon sources used in the amplifier must themselves be coupled into fibers. Hence, the overall detection efficiency required to close the detection loophole reduces to the product of the coupling efficiency $\eta_c$ by the detector efficiency $\eta_d$, but does not depend anymore on the transmission efficiency $\eta_t$.

We now perform a detailed analysis to assess the performance of our scheme where we consider two possibilities for the single-photon sources used in the amplifier: either on-demand or heralded sources. Note that the latter can be realized from a pair-source where the emission of an individual photon is heralded by the detection of the twin-photon, as implemented in Ref. \cite{Pittman04} from the parametric down conversion process. A single-photon source on-demand could then be obtained by adding a quantum memory. In the long run, on-demand sources based on quantum dots embedded in microcavities \cite{microcav} or single atoms inside high-finesse cavities \cite{atoms_cavity} are also potential candidates.

We consider the DIQKD protocol based on the CHSH inequality analyzed in \cite{QKD-DI-PRL}. Existing security proofs valid against collective attacks, assume perfect detectors \cite{QKD-DIFullLength,mckague}. We show in Appendix I, how to apply them to the case of imperfect devices and how to compute the corresponding key rate. Moving slightly away from a full device-independent scenario, we also consider the case where the end detectors are trusted and can be moved out of the black boxes. This means that the detectors are well characterized, have a known efficiency, and that the eavesdropper cannot tamper with them. In this case, a Bell violation can be observed independently of the detector efficiency $\eta_d$ and any local description is ruled out provided that the coupling $\eta_c$ of single-photons into optical fibers is high enough.
\begin{figure}[t]
{
\includegraphics[scale=0.35]{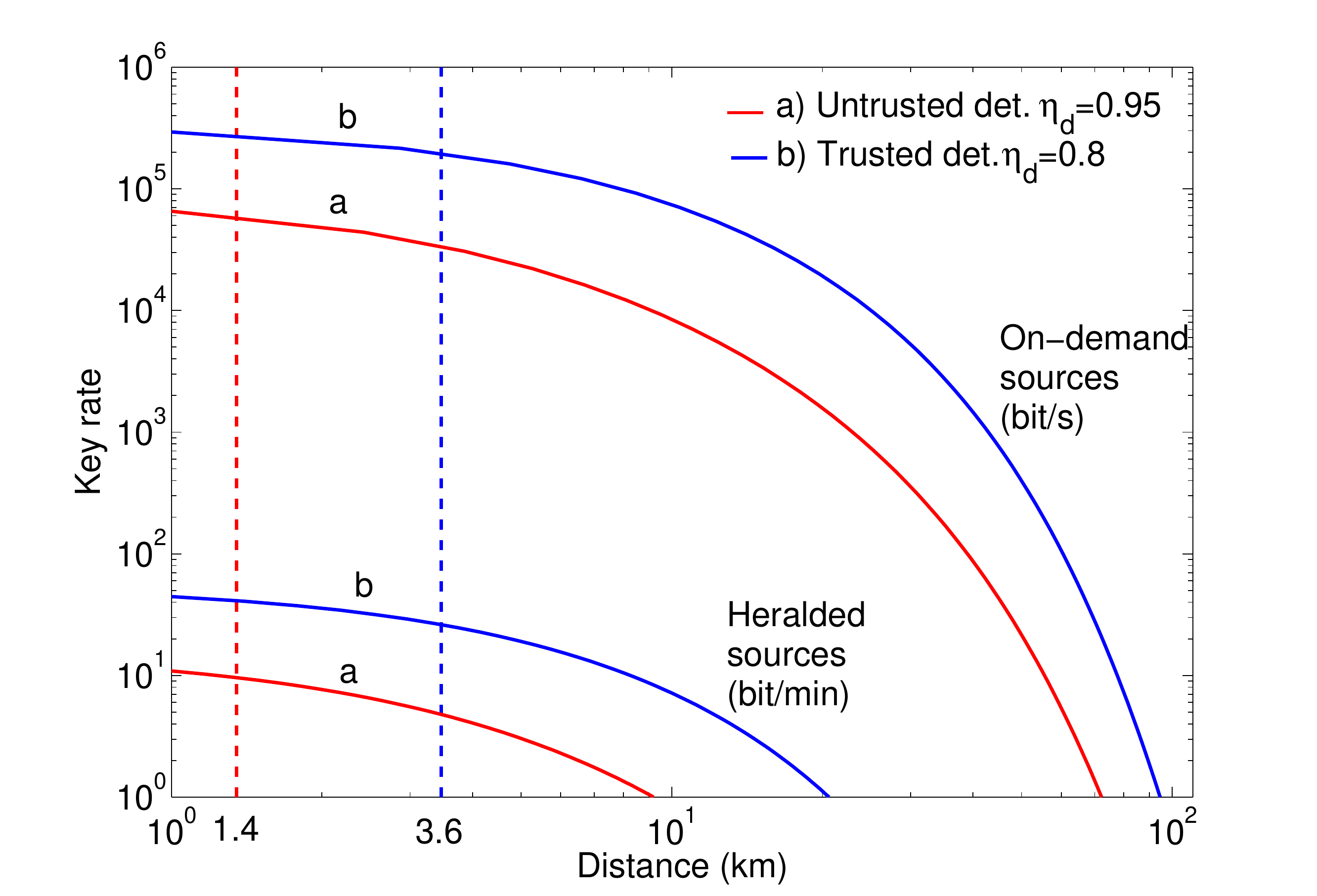}
 \caption{Key rate vs distance for DIQKD with imperfect devices (log-log scale). (Red) curves labelled a) correspond to untrusted detectors of efficiency $\eta_d=0.95$ (seen as part of the QKD black boxes);  (Blue)  curves labelled b) correspond to trusted detectors of efficiency $\eta_d=0.8$ (moved out of the QKD black boxes). The dotted vertical line represent the maximal distance above which no secret key can be extracted in the absence of an amplification process that counterbalances transmission losses. The two lower curves give the key rate (in bit/min) as a function of the distance for an amplifier based on heralded single-photon sources; the two upper curves  represent the key rate (in bit/s) for an amplifier with on-demand single-photon sources.}
\label{fig4}}
\end{figure}

To compute the key rate, we consider a fiber attenuation of $0.2$ dB/km, corresponding to telecom wavelength photons, and a coupling efficiency of $\eta_c=0.9$. The coupling efficiency of single-photons within optical fibers is being maximized in many laboratories and a coupling of 83\% was reported in \cite{Pittman04}. We assume that the photon-sources are excited with a repetition rate of $10$ Ghz \cite{10ghzsource}. We take all detectors to be photon-number resolving detectors with efficiency $\eta_d$ and we neglect dark counts. Note that superconducting transition-edge sensor detectors can already resolve telecom-wavelength photons and have 95\% efficiency with negligible noise \cite{Lita08}. Since we consider realistic sources, e.g. based on parametric down-conversion to provide high repetition rates, the dominant errors come from the multi-pair emissions which have to be made small by controlling the intensity of pumping lasers, i.e. the parameter $p$ for the entangled pair source and $p'$ for the pair-source used to produce heralded single-photons. For a given distance, we optimized the transmission coefficient $t$ and the pump dependent parameters $p$ and $p'$ to maximize the key rate, see Appendix II. The results of our calculations are presented in Figure~4 for untrusted detectors of efficiency $\eta_d=0.95$ and for trusted detectors of efficiency $\eta_d=0.8$.

In the absence of an amplification process, no secret key can be established beyond $1.4$km for untrusted detectors and beyond $3.6$km for trusted detectors. On the other hand, an implementation based on a qubit amplifier with heralded single-photon sources achieves rates of about $1$bit/min on distances of 10-20km and rates of about $1$bit/s on distances of 80-90km with on-demand single-photon sources. Note that contrarily to the situation without the amplifier, there is in principle no limitations other than technical ones
on these distances and they can be further extended, provided that one is willing to lower the key rate.

Finally, note that the physics behind the qubit-amplification is based on the bosonic character of indistinguishable photons. The temporal, spectral, spacial and polarization properties of modes produced by the entangled-pair source and by the single-photon sources (the modes $b$ and $c$ involved in the Bell measurement, see Figure \ref{fig3}) thus have to overlap. However, when the input state is an admixture between a qubit state and an empty component, as caused by losses, the optical path length does not require an interferometric control. The degree of indistinguishability of two photons is measured through the visibility $V$ of the ``Hong-Ou-Mandel" dip \cite{HOM}. Reference \cite{Pittman03} has reported a visibility $V=0.994$, largely sufficient for the successful implementation of our scheme (see analysis in Supplementary Informations II).

\paragraph{Conclusion.}
We have presented a simple qubit amplification scheme suited to the distribution of entanglement over large distances in a heralded way. This scheme could find applications, e.g., in traditional QKD \cite{lutke} or in the context of quantum repeaters \cite{qr}. Here, we show how to use it in DIQKD to overcome the problem of transmission losses.

An implementation of our proposal with heralded single-photon sources represent an experiment feasible with today's best technology that demonstrates DIQKD over 10-20 km of standard telecom fibers. The experiment promises to be difficult, though every single step of the proposed experiment has already been demonstrated. We see our proposal as a great challenge for the quantum communication community.

\paragraph{Acknowledgment.}
We thank H. Zbinden and one of the referees for pointing out simplifications in the implementation of the qubit amplifier. We also thank M. Afzelius, J.D. Bancal, N. Brunner, S. Massar, J. Min\'a\v{r}, H. de Riedmatten, P. Sekatski, C. Simon and R. Thew for valuable discussions.This work was supported by the ERC-AG QORE, the EU projects Qessence, the Swiss NCCR Quantum Photonics, and the Brussels-Capital region through a BB2B grant.

%%%%%%%%%%%%%%%%%%%%%%%%%%%%%%%%%%%%%%%%%%%%%%%%%%%%%

%%%%%%%%%%%%%%%%%%%%%%%%%%%%%%%%%%%%%%%%%%%%%%%%%%%%%

\section{Appendix I: DIQKD with photon losses}
Existing security proofs of DIQKD assume devices that always produce a conclusive answer, e.g., a $\pm 1$ result \cite{QKD-DI-PRLapp,QKD-DIFullLengthapp,mckagueapp}. Here we show how to apply them to the case of imperfect devices (including the transmission losses, the detector inefficiencies and the imperfect coupling of individual photons within optical fibers) and how to compute the corresponding key rate. We consider the DIQKD protocol based on the CHSH inequality introduced in \cite{ampapp} and first remind the known results in the case of lossless devices.

\subsection{Lossless devices}
In the ``2+3 bases" protocol introduced in \cite{ampapp}, Alice has three inputs $x=\{x_0,x_1,x_2\}$ and Bob two inputs $y=\{y_1,y_2\}$. All outputs $a=\{a_0,a_1,a_2\}$ and $b=\{b_1,b_2\}$ take binary values $a_0,a_1,a_2,b_1,b_2 \in \{-1,+1\}$. Most of the time, Alice and Bob use the inputs $x=x_0$ and $y=y_1$, and the raw key is extracted from the corresponding outputs $a_0$ and $b_1$. The amount of correlations between
Alice's and Bob's symbols is quantified by the quantum bit error rate (QBER) defined as
\begin{equation}\label{qber}
Q=P(a_0\neq b_1)\,.
\end{equation}
This parameter is related to the amount of classical communication needed for error correction. 

The inputs $x_1,x_2,y_1,y_2$ are used on a subset of the particles to bound the eavesdropper's information through the estimation of the CHSH quantity
\begin{equation}\label{chsh}
S=\langle a_1b_1\rangle+\langle a_1b_2\rangle+\langle a_2b_1\rangle-\langle a_2b_2\rangle\,,
\end{equation}
where the correlator $\langle a_ib_j\rangle$ is defined by $P(a_i=b_j)-P(a_i \neq b_j).$
The CHSH quantity bounds eavesdropper's information on the raw key and thus governs the privacy amplification process. Under collective attacks \cite{QKD-DI-PRLapp,QKD-DIFullLengthapp,mckagueapp}, eavesdropper's information is bounded by
\beq
I_E(S)\leq \chi(S)= h\left(\frac{1+\sqrt{(S/2)^2-1}}{2}\right)\,,
\eeq
where $h(x)=-x\log_2(x)-(1-x)\log_2(1-x)$ is the binary entropy. 

The achievable key rate $K$ after error-correction and privacy amplification is then given by 
\beq
K\geq 1-h(Q)-I_E(S)
\eeq

\subsection{Imperfect devices}\label{impdev}
If the devices of Alice and Bob have non-unit detection efficiency, the outputs $a$ and $b$ can take three values $\{i,\pm1\}$ where $i$ denotes an inconclusive result (the absence of a click). Let $\mu_{cc}$ denote the \emph{observed} probability of obtaining a conclusive result $(\pm 1)$ on each side; $\mu_{ci}$ of obtaining a conclusive result on Alice's side and an inconclusive one on Bob's side; and $\mu_{ic}$ of obtaining an inconclusive result on Alice's side and a conclusive one on Bob's side. Note that we assume for simplicity throughout the paper that the devices are such that $\mu_{cc}(xy)=\mu_{cc}$, $\mu_{ci}(xy)=\mu_{ci}$, and $\mu_{ic}(xy)=\mu_{ic}$ for all inputs $x,y$, i.e. the observed probabilities for the devices to produce a conclusive result are the same for all inputs. Our analysis, however, can be generalized in a straightforward way to more general cases.

We assume that the QBER and the CHSH value are computed as in Eqs. (\ref{qber}) and (\ref{chsh}) on the set of conclusive results $\pm 1$ using the renormalized probabilities $P(a_i,b_j)/\mu_{cc}$.

We consider the following possible strategies for the eavesdropper, Eve. Either she uses some quantum strategy $q$ that will produce two conclusive results on each side with certainty. This arises with probability $P_q$ and contributes by an amount $S_q$ to the CHSH violation. In this case, Eve's information is bounded by $I_E(q)\leq \chi(S_q)$ as determined in \cite{QKD-DIFullLengthapp,mckagueapp}. Or she uses a mixture $g$ of ``guessing" strategies where on each run at least one of the inputs $x$ of Alice or $y$ of Bob is assigned an inconclusive results. In this case, Eve may have full information, $I_E(g)\leq 1$, and the CHSH violation (given that two conclusive results have been obtained on each side) is $S_g\leq 4$.

Let $P_g$ be the proportion of events that arises from a mixture of guessing strategies and where both Alice and Bob obtain a conclusive result. We thus have $\mu_{cc}=P_q+P_g$. The observed CHSH violation is given by
\beq
S=\frac{P_qS_q+P_gS_g}{P_q+P_g}\leq \frac{P_qS_q+4P_g}{\mu_{cc}}
\eeq
and thus
\beq\label{sq}
S_q\geq \frac{\mu_{cc}S-4P_g}{\mu_{cc}-P_g}\,.
\eeq
The information of Eve is bounded by
\begin{eqnarray}\label{boundIa}
I_E&=&\frac{P_qI_E(q)+P_gI_E(g)}{P_q+P_g}\leq \frac{P_q\chi(S_q)+P_g}{\mu_{cc}}\nonumber\\
&\leq &\left[(\mu_{cc}-P_g)\chi(\frac{\mu_{cc}S-4P_g}{\mu_{cc}-P_g})+P_g\right]/\mu_{cc}\,,
\end{eqnarray}
where we have used (\ref{sq}) to obtain the last inequality. 
Now note that
\beq
P_g\leq \mu_{ci}+\mu_{ic}
\eeq
since every guessing strategy that contributes to $\mu_{cc}$ contributes at least with the same weight to either $\mu_{ci}$ or $\mu_{ic}$ (since the guessing strategies assign an inconclusive results to at least one of the inputs $x$ of Alice or $y$ of Bob). As the bound (\ref{boundIa}) is a monotonically increasing function of $P_g$, it is necessarily smaller than or equal to the solution with $P_g=\mu_{ci}+\mu_{ic}$. Writing
\beq\label{defmu}
\mu=\frac{\mu_{ci}+\mu_{ic}}{\mu_{cc}}
\eeq
we finally find that Eve's information is bounded by
\beq \label{boundI2a}
I_E(S,
\mu) \leq (1-\mu)\chi\left(\frac{S-4\mu}{1-\mu}\right)+\mu
\eeq
which only depends on the observable quantities $S, \mu_{cc}, \mu_{ci}, \mu_{ic}$. Putting all together, the key rate per conclusive event is given by $[1-h(Q)-I_E(S,\mu)]$ and the key rate per use of the device is then
\beq \label{keyruntr}
K\geq\mu_{cc}\left[1-h(Q)-I_E(S,\mu)\right]\,.
\eeq

As an illustration, consider the case where $\mu_{cc}=\eta^2$, $\mu_{ci}=\mu_{ic}=\eta(1-\eta)$, where $\eta$ is the detection efficiency of each box. The bound (\ref{boundI2a}) then becomes
\beq
I\leq \left[(3\eta-2)\chi\left(\frac{\eta S-8(1+\eta)}{3\eta-2}\right)+2(1-\eta)\right]/{\eta}\,.
\eeq
For the perfect singlet correlations satisfying $Q=0$, $S=2\sqrt{2}$, the key rate is positive as long as $\eta>2/(1+\sqrt{2})\simeq 0.8284$ which corresponds to the threshold required to close the detection loophole with the CHSH inequality.

\subsection{Imperfect devices with trusted detectors}\label{trustdet}
If the detectors can be trusted, we can move them out of boxes. Instead of assuming that the boxes have classical outputs $\pm 1$, we assume that they have two output channels, a ``$+1$" channel and a $``-1"$ channel, from each of which $N$ photons can be emitted. These photons may then produce clicks on trusted photon-number resolving detectors characterized by a known efficiency, $\eta_d$. The eavesdropper can control how many photons output the boxes, but he cannot control whether they will be detected or not by the trusted detectors.
The QKD black boxes are thus now characterized by the probabilities $\gamma_{jk,lm}$ that $j$ photons are emitted in Alice's $+1$ channel, $k$ in Alice's $-1$ channel, $l$ in Bob's $+1$ channel, and $m$ in Bob's $-1$ channel.

As a starting point, let us assume that Alice and Bob have photon-number resolving detectors with unit efficiency, $\eta_d=1$. In this case, Alice and Bob observe $(jk,lm)$ clicks in their detectors only if the boxes sent $(jk,lm)$ photons, i.e., Alice and Bob have direct information about the outcomes $(jk,lm)$ produced by the devices. We define the set of conclusive events as those where a unique photon is detected on each side. For instance, if Alice chooses the input $x_0$ and finds one photon in the $+1$ channel and no photon in the $-1$ channel, she associates to this event the output $a_0=+1$. If there is no photon in the $+1$ channel and one in the $-1$ channel, this corresponds to $a_0=-1$. All other possibilities, for instance no photons outputted by any channels or 2 photons outputted by one of the channels, are considered as inconclusive events, $a_0=i$.
As before the raw key and the parameters $Q$ and $S$ are defined on the subset of conclusive events.
The situation is then formally equivalent to the one discussed in the previous section where the box can either produce conclusive or inconclusive events, which are unambiguously recognized as such by Alice and Bob.  Eve's information is thus bounded by the expression (\ref{boundI2a}) for $I_E(S,\mu)$ and the key rate is given by (\ref{keyruntr}).

If, on the other hand, Alice and Bob have photon-number resolving detectors with a finite efficiency $\eta_d<1$, they can no longer determine unambiguously the outcomes $(jk,lm)$ generated by the devices.  For instance, if Alice obtains one click in the $+1$ channel, this could either correspond to the event $j=1$ where the unique produced photon has been detected or to an event $j=2$ where one of generated photons did not give a click.
The probabilities $\delta_{jk,lm}$ that Alice and Bob obtain $(jk,lm)$ clicks can be easily computed from the probabilities $\gamma_{j'k',l'm'}$ that the boxes emit $(j'k',l'm')$ photons using the fact that the probability to detect $n$ photons when $n'$ have been produces in a given channel is $p_{nn'}={n'\choose n}\eta_d^{n}(1-\eta_d)^{n'-n}$.

As before, we define the set of conclusive events as those where only one photon is \emph{detected} and we denote by $\mu_{cc}$ the probability of finding two conclusive results on each side. We also introduce the notation 
\begin{equation}\label{mu1}
\tilde\mu_{cc}=\gamma_{10,10}+\gamma_{10,01}+\gamma_{01,10}+\gamma_{01,01}
\end{equation} for the probability that the devices  \emph{produce} a single photon on each side. This quantity is identical to $\mu_{cc}$ when the detectors have unit efficiency $\eta_d=1$. Similarly, we introduce the notation
\begin{equation}\label{mu2}
\tilde\mu_{ci}=\sum_{j+k\neq 1} \gamma_{10,jk}+\gamma_{01,jk}
\end{equation}
for the probability that the devices output a single photon on Alice's side and strictly more or strictly less than one photon on Bob's side, and define analogously $\tilde\mu_{ic}$. Again, these quantities reduce to $\mu_{ci}$ and $\mu_{ic}$, respectively, when the detectors have unit efficiency, $\eta_d=1$.

A fraction $\eta_d^2\tilde{\mu}_{cc}/\mu_{cc}$ of the raw key originates from events where a single photon has been produced by the boxes and has successfully been detected by Alice's and Bob's detector. This part of the raw key corresponds to the ideal situation discussed above when $\eta_d=1$.  The information that Eve has on it, is thus bounded by $I_E(\tilde S,\tilde\mu)$, where $\tilde S$ is the CHSH violation conditional to the emission of one photon from each boxes and $\tilde \mu=(\tilde{\mu_{ci}}+\tilde{\mu_{ic}})/\tilde{\mu_{cc}}$ is defined through the relations (\ref{mu1}) and (\ref{mu2})~\footnote{Indeed, $I(S,\mu)$ is an intrinsic property of the devices; it represents an information theoretic bound on the information of Eve when the boxes output a single photon.}.
Note that even though Alice and Bob cannot know for each individual run exactly how many photons where sent by the boxes, i.e., which outcome $(jk,lm)$ was precisely produced, they can nevertheless determine the probabilities $\gamma=\{\gamma_{jk,lm}\}$ characterizing the output of the devices from their observed detection statistics $\delta=\{\delta_{jk,lm}\}$. This information is sufficient to determine in turn the parameters $\tilde S$ and $\tilde \mu$ above and thus to compute $I_E(\tilde S,\tilde \mu)$.

The remaining fraction $1-\eta_d^2\tilde\mu_{cc}/\mu_{cc}$ of the raw key originates from events where more than one photon are produced at each side, but only one was detected. In this case, we conservatively give to Eve all information about the outcomes. In total, Eve information is thus equal to
\beq\label{keyrtrust2}
I_E(\delta)\leq \frac{\eta_d^2\tilde\mu_{cc}}{\mu_{cc}} I_E(\tilde S,\tilde\mu) + \left(1-\frac{\eta_d^2\tilde\mu_{cc}}{\mu_{cc}}\right)
\eeq
which can be determined solely from the observed statistics $\delta$. 
 Finally, taking into account the probability with which Alice and Bob observe a conclusive results, we find that the key rate is given by
\beq
K\geq \mu_{cc}\left[1-h(Q)-I_E(\delta)\right]\,.
\eeq

\section{Appendix II: Key rate with a qubit amplifier implementation}

Here, we detail the calculation of the achievable key rate when DIQKD is implemented with a heralded qubit amplifier. \\

\subsection{Single-photon sources}

The key rates have been estimated by considering various resources. In particular, the single-photon sources required within the qubit amplifier can be either heralded or on-demand. 

Let us first focus on the implementation of a heralded single-photon source from a pair source based on the parametric down convertion (PDC) process. The state resulting from the PDC process is well approximated by 
\begin{equation}
|00\rangle\langle00|+p'|11\rangle\langle11|+p'^2|22\rangle\langle22| + o(p'^3).
\end{equation} 
$p'$ is the probability for the successful emission of one pair. The first (second) Fock state gives the number of photons in the signal (idler) mode. The detection of one member of a pair can then be used to herald the presence of the other. This provides a heralded single-photon source as required in the proposed qubit-amplifier. The probability for a detector that is photon number resolving, but that has non-unit efficiency $\eta_d,$ to detect a single photon in a predetermined mode, given that there are $n$ photons present in that mode, is $n\eta_d(1-\eta_d)^{n-1}.$ Hence, the state conditional on the detection of a single-photon is  
\begin{equation}
|1\rangle\langle1|+2p'(1-\eta_d)|2\rangle\langle2|+o(p'^2)
\end{equation}
and the success probability for the heralding signal is $P_S=p'\eta_d.$

Our proposal requires two heralded photons with orthogonal polarizations. In what follows, we focus on an implementation based on two separate crystals. Note that two heralded photons could be produced by a single non-linear crystal by selecting only the emissions of double pairs of photons. In other terms, the detection of two photons with orthogonal polarization in e.g. the ``idler" mode heralds the production of the two desired photons in the signal mode.\\

A single-photon source on-demand is more difficult to implement in practice but it provides a higher key rate than the one from a heralded single-photon source. For the first demonstration experiments, the most promising approach may be the use of heralded single-photon source based on parametric down-conversion, as previously described, combined with a quantum memory for light. In the long run, sources
based on quantum dots \cite{dotsapp} embedded in microcavities \cite{microcavapp} are likely to offer higher repetition rates. Single atoms inside high-finesse cavities \cite{atoms_cavityapp} are also potential candidates. In our paper, the state generated by on-demand sources is assumed to be described by the Fock state $|1\rangle$ and to be produced with the probability $P_S=1.$
  
\subsection{State conditional to a successful amplification}

\begin{figure}
{\includegraphics[scale=0.65]{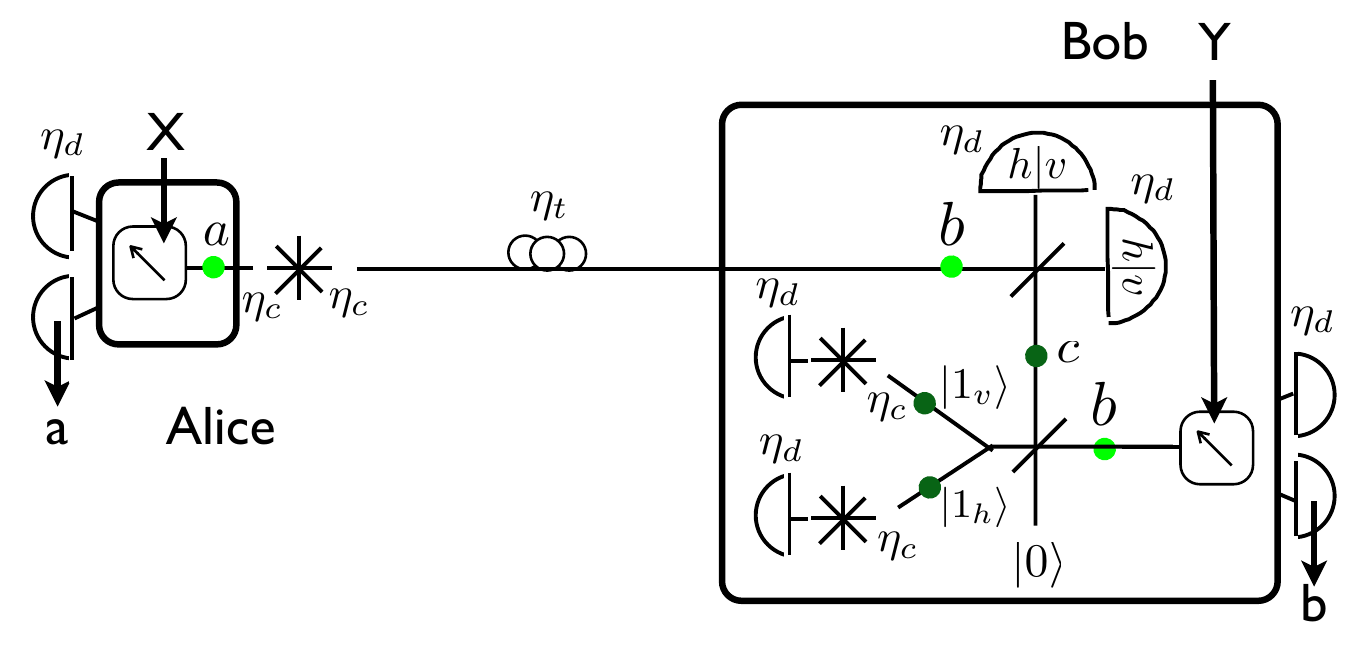} \caption{Proposed setup for the implementation of device-independent quantum key distribution based on heralded qubit amplification. Here, the photon sources are based on the parametric down conversion process. The entangled-photon source is located close to Alice's location. Two sources of heralded single-photons are located within the qubit amplifier at Bob's location. 
The photons are coupled into optical fibers with efficiency $\eta_c.$ The detectors have non-unit efficiencies given by $\eta_d.$ The efficiency of the transmission line is labelled $\eta_t.$}
\label{fig5}}
\end{figure}

The evaluation of the achievable ket rate is based on the knowledge of the state that is shared by Alice and by Bob. In the ideal case where we take into account the transmission losses only, the state resulting from a successful amplification is given by the equation (2) of the main text. However, to properly assess the performance of our scheme, we take into account other imperfections in what follows. (Note that these imperfections are represented schematically in Figure \ref{fig5}.) 

The starting point is the initial state which is the product of three states. The first one is associated to the stochastic emission of entangled photon-pairs at Alice's location where we now explicitly write the $O(p^2)$ terms corresponding to double-pair emissions. The two others are associated to the single-photon emissions within the qubit amplifier, one horizontally polarized and one vertically polarized 
\begin{eqnarray}
\nonumber
&&\Big[|1_h\rangle\langle1_h|+2p'(1-\eta_d)|2_h\rangle\langle2_h|\Big] \otimes \\
&&\nonumber \Big[|0\rangle\langle0|+p|\frac{1}{\sqrt{2}}\left(a_h^\dagger b_h^\dagger+a_v^\dagger b_v^\dagger\right)\rangle\langle \frac{1}{\sqrt{2}}\left(a_h^\dagger b_h^\dagger+a_v^\dagger b_v^\dagger\right)| +\\
\nonumber
&& \frac{3}{4}p^2 |\frac{1}{2\sqrt{3}}\left(a_h^\dagger b_h^\dagger+a_v^\dagger b_v^\dagger\right)^2\rangle\langle \frac{1}{2\sqrt{3}}\left(a_h^\dagger b_h^\dagger+a_v^\dagger b_v^\dagger\right)^2|\Big] \otimes \\
&& \nonumber \Big[|1_v\rangle\langle1_v|+2p'(1-\eta_d)|2_v\rangle\langle2_v|\Big].
\end{eqnarray}
(If the single-photons are produced on-demand, the initial state is similar but with $p'=0.$) The modes produced by the entangled-pair source, labelled $a$ and $b,$ are each coupled into optical fibers with efficiency $\eta_c.$ Alice performs measurement on the mode $a.$ The mode $b$ is sent to Bob's location using an optical fiber with the transmission efficiency $\eta_t.$ The photons emitted by the heralded sources located within the qubit amplifier, are coupled into optical fibers with efficiency $\eta_c.$ Then, they are sent through a partial beamsplitter with transmission $t$ to form an entangled state involving the modes $c_h,$ $c_v,$ and the modes later on detected by Bob. The modes $b$ and $c$ are combined on a 50/50 beamsplitter to perform a partial Bell state measurement based on photon detectors with non-unit efficiency $\eta_d.$ 

From a perturbative calculation, i.e. keeping only the terms at the order $o(p^2),$ $o(p'^2)$ and $o(pp'),$ we derived explicitly the state $\rho$ resulting from a successful Bell measurement~\footnote{Note that we checked that the optimal values for $p$ and $p'$ are such that the terms that we leave at the order $o(p^3),$ $o(p'^3),$ $o(p^2 p')$ and $o(p p'^2)$ are negligible.} . This mixed state has many components
\begin{eqnarray}
\rho&=& \tilde P_{00}\, \rho_{00} + \tilde P_{01}\,\rho_{01} +\tilde P_{10}\,\rho_{10}\nonumber\\
&&\quad +\tilde P_{11}\,\rho_{11}+\tilde P_{02}\,\rho_{02}+\tilde P_{20}\,\rho_{20}\\
&&\quad +\tilde P_{12}\,\rho_{12}+\tilde P_{21}\,\rho_{21}+\tilde P_{22}\,\rho_{22}\nonumber
\end{eqnarray}
corresponding to different cases where Alice and Bob get each either zero, one or two photons. By summing the weights $\tilde{P}_{ij}$ of these components, one obtained the success probability for the heralded qubit amplification $P_H.$ The renormalized weights $P_{ij}=\tilde P_{ij}/P_H$ correspond to the probabilities that Alice gets $i$ photons and that Bob gets $j$ photons exactly, knowing that the amplification succeeded.

\subsection{Key rates for imperfect devices}

To compute the key rate, we consider a particular implementation of the ``2+3 bases" protocol \cite{ampapp} where Alice chooses to apply one out of three possible measurements $x_0=\sigma_z,$ $x_1=(\sigma_z+\sigma_x)/\sqrt{2},$ $x_2=(\sigma_z-\sigma_x)/\sqrt{2}$ and where Bob chooses one measurement out of two, either $y_1=\sigma_z$ or $y_2=\sigma_x$. This specific choice maximizes the CHSH polynomial when Alice and Bob share a maximally entangled state.

\subsubsection{Untrusted detectors}
We first consider the case where Alice's and Bob's detectors are untrusted and are thus a part of the QKD black-boxes. We follow the analysis reported in subsection \ref{impdev} to obtain the secret key rate. We remind that a conclusive event corresponds to a single detector click. The probabilities $\mu_{cc},$ $\mu_{ci}$ and $\mu_{ic}$ to obtain conclusive-conclusive, conclusive-inconclusive, and inconclusive-conclusive events, respectively, are functions of the probabilities $P_{ij}$ to have $i$ photons on Alice's side and $j$ photon on Bob's side, defined by
\begin{subequations}
\begin{eqnarray}
\label{mucc}
\nonumber
\mu_{cc}&=&\eta_d^2 P_{11}+2(1-\eta_d)\eta_d^2(P_{21}+P_{12})\\
&+&4\eta_d^2(1-\eta_d)^2P_{22},
\end{eqnarray}
\begin{eqnarray}
\nonumber
\mu_{ci}&=&\eta_dP_{10}+\eta_d(1-\eta_d)P_{11}+2\eta_d(1-\eta_d)^2P_{21}\\
\nonumber
&+&\left(\eta_d^3+\eta_d\left(1-\eta_d\right)^2\right)P_{12}+2\eta_d(1-\eta_d)P_{20}\\
\label{muci}
&+&\left(2\eta_d^3(1-\eta_d)+2\eta_d(1-\eta_d)^3\right)P_{22}
\end{eqnarray}
\begin{eqnarray}
\nonumber
\mu_{ic}&=&\eta_dP_{01}+\eta_d(1-\eta_d)P_{11}+2\eta_d(1-\eta_d)^2P_{12}\\
\nonumber
&+&\left(\eta_d^3+\eta_d\left(1-\eta_d\right)^2\right)P_{21}+2\eta_d(1-\eta_d)P_{02}\\
\label{muic}
&+&\left(2\eta_d^3(1-\eta_d)+2\eta_d(1-\eta_d)^3\right)P_{22}.
\end{eqnarray}
\end{subequations}
The QBER $Q$ and the CHSH value $S$ are given by
\beq\label{qber2}
Q=\eta_d^2 Q_{11} + 2(1-\eta_d)\eta_d^2(Q_{21}+Q_{12})+4\eta_d^2(1-\eta_d)^2 Q_{22}
\eeq
and
\beq\label{chsh2}
S=\eta_d^2 S_{11} + 2(1-\eta_d)\eta_d^2(S_{21}+S_{12})+4\eta_d^2(1-\eta_d)^2 S_{22}
\eeq
where $Q_{ij}$ and $S_{ij}$ represent the QBER and the CHSH values computed on the state $P_{ij}\rho_{ij}/\mu_{cc}$ for the measurement settings $x$ and $y$ specified above. 
The key rate per conclusive event, given by $[1-h(Q)-I_E(S,\mu)]$ is obtained from Eve's information $I_E(S,\mu)$ which is calculated from Eq. (\ref{boundI2a}). Taking the success probability for the single-photon emission $P_S$ and the success probability for the qubit amplification $P_H$ into account, we deduce the key rate per second
\begin{eqnarray}
\label{keyrateuntrusted}
K &= &r\times P_S^2 P_H \times \\
\nonumber
 && \mu_{cc}\left(1-h(Q)- \left((1-\mu)\chi\left(\frac{S-4\mu}{1-\mu}\right)+\mu\right)\right)
\end{eqnarray}
where $r$ is the repetition rate of sources. Let us remind that when the qubit amplifier uses on-demand single-photon sources, $P_S=1.$

\subsubsection{Trusted detectors}

We now consider the case where the detectors are trusted and moved out of the boxes. This corresponds to the analysis developed in subsection \ref{trustdet}. The raw key is formed as before on the set of conclusive events corresponding to a single detector click on each side. The parameter $\mu_{cc}$ and the QBER $Q$ are thus given by (\ref{mucc}) and  (\ref{qber2}) as before. However, the CHSH value $\tilde{S}$ is now calculated on the state $\rho_{11}$ corresponding to a single photon on each side. Finally, $\tilde \mu$ is defined from the following parameters
\begin{eqnarray}
\nonumber
&&\tilde{\mu}_{cc}=P_{11}, \\
\nonumber
&& \tilde{\mu}_{ci}=P_{10}+P_{12}, \\
\nonumber
&&\tilde{\mu}_{ic}=P_{01}+P_{21}.
\end{eqnarray}
The key rate per conclusive event is then given by $[1-h(Q)-I_E(\delta)]$ where Eve's information $I_E(\delta)$ is calculated from Eq. (\ref{keyrtrust2}), and the key rate per second is given by
\begin{equation}
\label{keyrsecondt}
K = r P_S^2 P_H \mu_{cc}\left(1-h(Q)- I_E(\delta))\right)
\end{equation}
where $r$ is the repetition rate of sources, $P_S$ is the success probability for the single-photon emission and $P_H$ is the success probability for the qubit amplification. Let us remind again that when the qubit amplifier uses on-demand sources, $P_S=1.$

\subsection{Requirements on the overall detection efficiency}

Let us roughly estimate the overall detection efficiency which is required to rule out attacks based on the detection loophole. If Bob's box contains a qubit amplifier and if the sources are weakly excited such that the double pair emissions can be neglected, we have $P_{02} \approx P_{20} \approx P_{12} \approx P_{21} \approx P_{22} \approx 0.$ Furthermore, if the reflectivity of the beam-splitter located within the amplifier is weak enough, the conditional probability to distribute one photon at a given location is mainly determined by the coupling of single photons within an optical fiber $\eta_c,$ i.e. $P_{11}=\eta_c^2$ and $P_{10}+P_{01}=2\eta_c(1-\eta_c).$ Note that in this case, the state corresponding to the event where one photon is generated at each side $\rho_{11}$ is a pure maximally entangled state $\rho_{11}=\phi_+=1/\sqrt{2}(a_hb_h+a_vb_v).$ \\

First, consider the case where the detectors are untrusted. The QBER, calculated from Eq. (\ref{qber2}), reduces to zero since the correlations observed from a maximally entangled state are perfect. Eve's information $I_E(S,\mu)$ is obtained from Eq. (\ref{boundI2a}) where the CHSH value, calculated from Eq. (\ref{chsh2}), is given by $S=2\sqrt{2}$ and where the parameter $\mu$ (see Eq. (\ref{defmu})) is defined as $\mu=2(1-\eta_d\eta_c)/(\eta_d\eta_c).$ One concludes that the difference of mutual informations $1-h(Q)-I_{E}(S,\mu)$ is positive as long as the argument of $\chi$ is greater than 2 (see Eq. (\ref{boundI2a})), i.e. if $\eta_d\eta_c \geq 2/(1+\sqrt{2})\simeq 0.8284.$ Thus, the proposed protocol requires a minimum value for the product of the detector efficiency by the coupling efficiency $(\eta_d\eta_c)^{min} = 0.8284$ which corresponds to the threshold required to close the detection loophole with the CHSH inequality. If this is satisfied, the distribution of a quantum key is made possible, independently of the proper functioning of the devices, for arbitrary long distances. Note that without qubit amplifier, $\eta_d \eta_c$ has to be replaced by the product $\sqrt{\eta_t} \eta_d \eta_c$ where $\eta_t$ is the transmission efficiency of the optical fiber connecting Alice's and Bob's location. This means that for $\eta_d=0.95$ and $\eta_c=0.9,$ DIQKD is possible only for distances smaller than $1.4$ km.\\

In case where the detectors can be trusted, the QBER is unchanged and hold to zero but eavesdropper's information $I_E(\delta)$ has to be calculated from Eq. (\ref{keyrtrust2}) which reduces to $I_E(\tilde{S},\tilde{\mu})$ since $\mu_{cc}=\eta_d^2 \tilde{\mu}_{cc}.$ We find $\tilde{S}=2\sqrt{2}$ and $\tilde{\mu}=2(1-\eta_c)/\eta_c$ so that any attacks based on the detection loophole is ruled out as long as the coupling efficiency is greater than $\eta_c^{min} = 0.8284$ independently of the detection efficiency. We emphasize that without qubit amplifier, $\eta_c$ has to be replaced by the product $\sqrt{\eta_t} \eta_c$ and for $\eta_c=0.9,$ Alice and Bob cannot exchange a secure key if the distance separating them is larger than $3.6$ km.\\

\subsection{Performance of the proposed protocol}

Finally, let us detail how the achievable key rate is evaluated, first by focusing on the case where the detectors are untrusted. We fix the detector efficiency $\eta_d$ and the coupling efficiency $\eta_c.$ For a given distance, we optimize the success probability for the entangled-pair emission $p$ and for the single-photon emission $p',$ as well as the transmission of beam splitter located within the qubit amplifier $t$ in order to maximize the key rate (\ref{keyrateuntrusted}). 

Assume for concreteness $\eta_c=0.9$ and $\eta_d=0.95.$ For $10$ km, we found that the optimal success probability for entangled-pair emission and for single-photon emission are respectively $p=2\times 10^{-3}$ and $p'=3\times 10^{-3}$ and that the optimal transmission of the beam splitter located within the qubit amplifier is $t \approx 0.98.$ If the sources are excited with a repetition rate of 10 Ghz, this leads to a key rate of roughly $1$ bit/min. If the required single-photons can be produced on-demand, one has $P_S=1$ and DIQKD can be performed over $90$ km with a key rate of 0.1 bit/s. 

In case where the detectors are trusted, one has to maximize the formula (\ref{keyrsecondt}). 
With $\eta_c=0.9,$ $\eta_d=0.8$ and for $10$ km, the optimal values are $p=7\times 10^{-3}$ and $p'=4\times 10^{-3}$ and $t \approx 0.97.$ For a repetition rate of 10 Ghz, the achievable key rate is roughly of $7$ bits/min. If the single-photons are on-demand, a secret key can be transmitted over $90$ km with a rate of 2 bits/s. \\

\begin{figure}[t]
{
\includegraphics[scale=0.33]{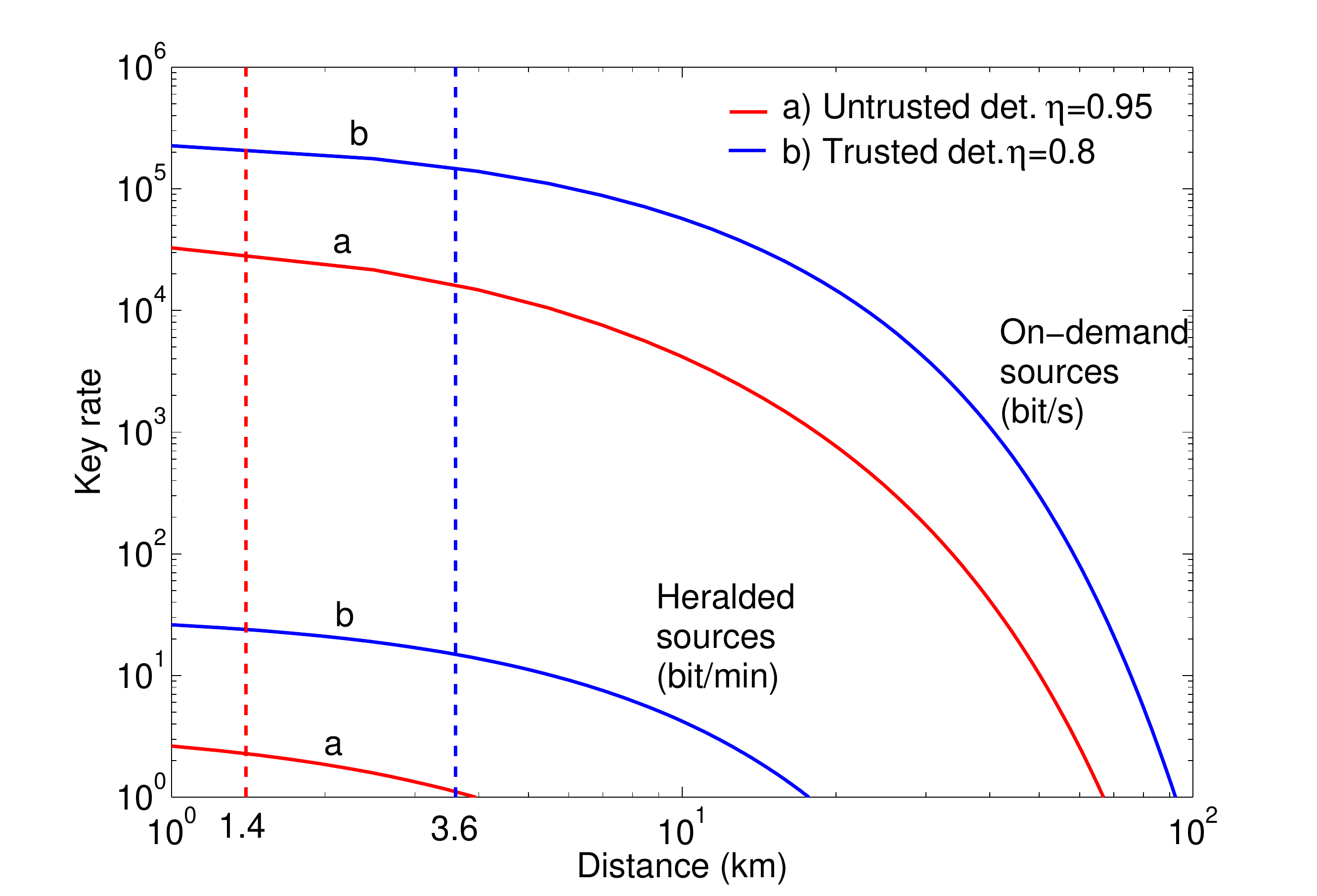}
 \caption{Key rate vs distance for DIQKD (log-log scale) when the mode overlap is imperfect (V=0.994). (Red) curves labelled a) correspond to untrusted detectors of efficiency $\eta_d=0.95$;  (Blue) curves labelled b) correspond to trusted detectors of efficiency $\eta_d=0.8$. The dotted vertical line represent the maximal distance above which no secret key can be extracted without qubit amplification. The two lower curves give the key rate (in bit/min) as a function of the distance for an amplifier based on heralded single-photon sources; the two upper curves represent the key rate (in bit/s) for an amplifier with on-demand single-photon sources.}
\label{fig6}}
\end{figure}

\subsection{Photon indistinguishability}

To herald the remote distribution of entanglement using the qubit amplifier, the photons involved in the Bell measurement have to be indistinguishable. The degree of indistinguishability of two photons can be quantified by their mode overlap which corresponds experimentally to the visibility of the ``Hong-Ou-Mandel" (HOM) dip \cite{HOMapp}. We have estimated with a simple model based on discrete modes that partial overlaps lead to a phase noise on the heralded entanglement. Alice and Bob do not share the state $\phi_+$ anymore but instead a state which has an admixture of $\phi_-,$ i.e. 
$
F|\phi_+\rangle\langle\phi_+|+(1-F)|\phi_-\rangle\langle\phi_-|
$
where $F=(1+V^3)/2.$ $V$ is the HOM dip visibility which is supposed to be the same for the modes $\{b, c_h\},$ $\{b, c_v\},$ or $\{c_h, c_v\}.$ This phase noise reduces potentially the CHSH violation and thus the key rate. In Figure \ref{fig6}, we present the achievable key rate as a function of the distance for $V=0.994$ corresponding to the visibility reported in Ref. \cite{Pittman03app}. One sees in comparison with the result of the Figure 4 (main text) that the key rate is divided by a factor of 3.5 in the case of untrusted detectors with heralded sources but is essentially unchanged in the case of trusted detectors with on-demand sources. 
In conclusion, small imperfections in the mode overlap do not dramatically change the performance of our protocol. 

\subsection{Remarks on the quantum relays for DIQKD}

Note that a standard quantum relay made with SPDC sources is not an alternative solution for the implementation of device independent quantum key distribution. The central problem with a quantum relay comes from the multi-pair emission and this problem cannot be circumvented only by reducing the power of the pump injected in the nonlinear crystal. To mitigate the multi-pair problem in a standard quantum relay, it is necessary to post-select either the events where there is one detection at both ends of the chain (at each Bob's and Alice's locations) or the events where a predetermined detector located at one end of the chain (either at Alice's or Bob's location) clicks provided that the closest source is very weakly excited. Such a postselection is fine for standard QKD. However, postselections of that kind is incompatible with DIQKD since they open inevitably the detection loophole. Our proposal fundamentally differs from a conventional quantum relay. It allows to distribute high-quality entanglement provided that the two photons required within our qubit amplifier are produced in a heralded way.x\\

%%%%%%%%%%%%%%%%%%%%%%%%%%%%%%%%%%%%%%%%%%%%%%%%%%%%%

\end{document}